\documentclass[fleqn,10pt]{wlscirep}
\pdfoutput=1
\usepackage{subfigure}
\usepackage{comment}
\usepackage[utf8]{inputenc}
\usepackage[T1]{fontenc}
\usepackage{lineno}
\usepackage{hyperref}

\title{Large-scale Multi-layer Academic Networks Derived from Statistical Publications}

\author[1]{Tianchen Gao}
\author[1]{Yan Zhang}
\author[2,*]{Rui Pan}
\author[3]{Hansheng Wang}
\affil[1]{Xiamen University School of Economics, Xiamen, 361005, China}
\affil[2]{Central University of Finance and Economics School of Statistics and Mathematics, Beijing, 100081, China}
\affil[3]{Peking University Guanghua School of Management, Beijing, 100871, China}

\affil[*]{corresponding author(s): Rui Pan (panrui\_cufe@126.com)}

\begin{abstract}
The utilization of multi-layer network structures now enables the explanation of complex systems in nature from multiple perspectives. Multi-layer academic networks capture diverse relationships among academic entities, facilitating the study of academic development and the prediction of future directions. However, there are currently few academic network datasets that simultaneously consider multi-layer academic networks; often, they only include a single layer. In this study, we provide a large-scale multi-layer academic network dataset, namely, LMANStat, which includes collaboration, co-institution, citation, co-citation, journal citation, author citation, author-paper and keyword co-occurrence networks. Furthermore, each layer of the multi-layer academic network is dynamic. Additionally, we expand the attributes of nodes, such as authors' research interests, productivity, region and institution. Supported by this dataset, it is possible to study the development and evolution of statistical disciplines from multiple perspectives. This dataset also provides fertile ground for studying complex systems with multi-layer structures. 
\end{abstract}
\begin{document}

\flushbottom
\maketitle

\thispagestyle{empty}


\section*{Background \& Summary}

Relational structures consisting of different types of interactions among several groups of entities are very common nowadays. As a useful tool for analyzing this type of data, multi-layer networks have gained increasing attention in recent years due to their ability to capture the complexity of real-world systems \cite{kivela2014multilayer}. Multi-layer academic networks are a specific type of multi-layer network that consist of multiple layers of relationships among academic entities, such as researchers, institutions, papers or journals. Each layer represents a different type of relationship, and these layers can be analyzed separately to capture specific information for each layer, or in combination to leverage information that may be shared across different relations.\cite{zhang2020flexible} Therefore, it is necessary to analyze multi-layer networks from multiple perspectives to gain a more comprehensive understanding of the underlying system.
 
Typical examples of multi-layer academic networks include the collaboration network that represents co-authorship relationships among researchers \cite{newman2001scientific},  the citation network that represents citation relationships among papers \cite{newman2008physics}, and the journal citation networks that represent citation relationships among journals \cite{su2011prestigerank}. These networks have been utilized in various disciplines, such as computer science \cite{zhou2018academic}, medicine \cite{peng2022bibliometric}, physics \cite{zhao2022utilizing}, sociology \cite{shiau2018examining} and others. They have been used for various purposes, such as identifying research areas \cite{ji2022co}, evaluating research impact \cite{saari2011developmental}, predicting scientific trends \cite{zhang2021measuring}, studying the diffusion of scientific knowledge \cite{uzzi2013atypical}, and supporting science policy and decision-making \cite{meng2016collaboration}. Overall, multi-layer academic networks provide a powerful tool for understanding and analyzing the complex relationships that underlie academic communities and their impact on scientific knowledge production and dissemination \cite{xu2023covariate}.

Despite their usefulness, most of the available data on academic networks generally only consist of one single layer, with collaboration and citation networks being the most widely studied types \cite{newman2001structure,newman2004coauthorship,chen2010community}. To illustrate large-scale academic networks, we provide some specific examples. One such example is the Arxiv HEP-PH citation network, which focuses on high-energy physics phenomenology. The network is built from the e-print arXiv and includes citations among 34,546 papers published between 1993 and 2003, with 421,578 edges linking them \cite{leskovec2005graphs}. Another example is the Arxiv COND-MAT collaboration network, which also originates from the e-print arXiv repository. It captures scientific collaborations among authors who submitted papers to the condensed matter physics section between 1993 and 2003. This network includes 23,133 nodes and 93,497 edges, providing a comprehensive view of the COND-MAT section's history \cite{leskovec2005graphs}. 
However, publicly available multi-layer academic networks related to statistics are scarce. Recently, co-citation and co-authorship networks of statisticians are studied utilizing a dataset consisting of 83,331 articles published in 36 representative journals in statistics, probability, and machine learning, spanning from 1975 to 2015\cite{ji2022co}.

In this work, we collect data from 42 statistical journals published between 1981 and 2021 from the Web of Science ({\it www.webofscience.com}). Our LMANStat dataset includes basic information on 97,436 papers, including their title, abstract, keywords, publisher, published date, volume and pages, document type, citation counts, author information (name, ORCID, address, region, and institution), as well as their reference lists. An example of such a paper is listed in Table \ref{tab:data_example}. Based on this information, we construct multi-layer academic networks, including collaboration network, co-institution network, citation network, co-citation network, journal citation network, author citation network, author-paper network, and keyword co-occurrence network. These networks change dynamically over time, providing a dynamic analytical perspective during analysis. Moreover, we also include rich nodal attributes of authors, such as the authors' research interests, to enhance the usefulness of our dataset.

\begin{table}[ht]
\centering
\newcommand{\tabincell}[2]{\begin{tabular}{@{}#1@{}}#2\end{tabular}}
\begin{tabular}{ll}
    \hline
    \textbf{Variable} & \textbf{Example} \\
    \hline
    Title & Regression shrinkage and selection via the Lasso\\
    Abstract &  \tabincell{l}{We propose a new method for estimation in linear models. The `lasso' minimizes the residual\\ sum of squares subject to the sum of the absolute value of the coefficients being less than a \\constant. Because of the nature of this constraint it tends to produce some coefficients that are\\ exactly 0 and hence gives interpretable models...}\\
    Keywords & Quadratic programming, Regression,  Shrinkage, Subset selection\\
    Publisher & Journal of the Royal Statistical Society Series B: Statistical Methodology\\
    Published date & 1996\\
    Volume and pages & Volume: 58; Pages: 267-288 \\
    Document type & Article \\
    Citation counts & 21,383 (until 2022)\\
    Author information & Tibshirani, R@University of Toronto\\
    Reference list &  \tabincell{l}{$\left[1\right]$ Linear-model selection by cross-validation \\Shao, J@1993@Journal of the American Statistical Association\\ $\left[2\right]$ Multivariate adaptive regression splines\\Friedman, J@1991@Annals of Statistics \\$\left[3\right]$ ...}\\
	
    \hline
    \end{tabular}
    \caption{\label{tab:data_example} An example of a published paper in the LMANStat dataset. Basic information such as title, abstract, keywords, citation counts, author information, and reference list can be obtained. The citation counts are provided by the Web of Science.}
\end{table}

\section*{Methods}

\begin{figure}[!ht]
	\centering
	\includegraphics[width=0.65\textwidth]{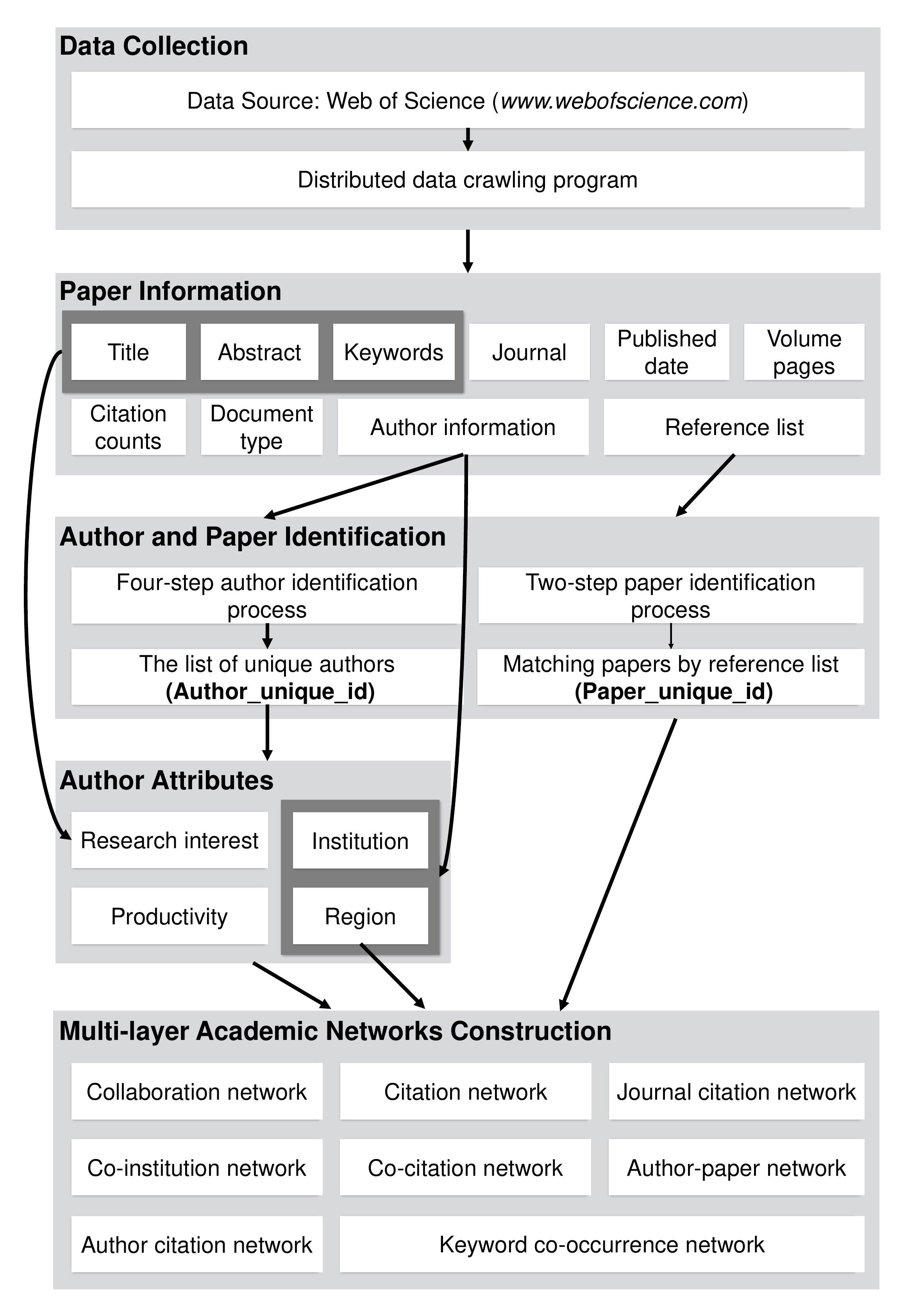}
        \caption{A schematic overview of the steps involved in our methodology.}
	\label{fig:data_collection}
\end{figure}

In this section, we present a comprehensive overview of our methodology, which covers the complete workflow from data collection to data cleaning, as well as the construction of multi-layer academic networks. The process is visually depicted in Figure \ref{fig:data_collection}. Subsequently, we provide detailed explanations regarding author and paper identification, the extraction of author attributes, and the construction of multi-layer academic networks.

\begin{figure}[!ht]
	\centering
	\includegraphics[width=0.99\textwidth]{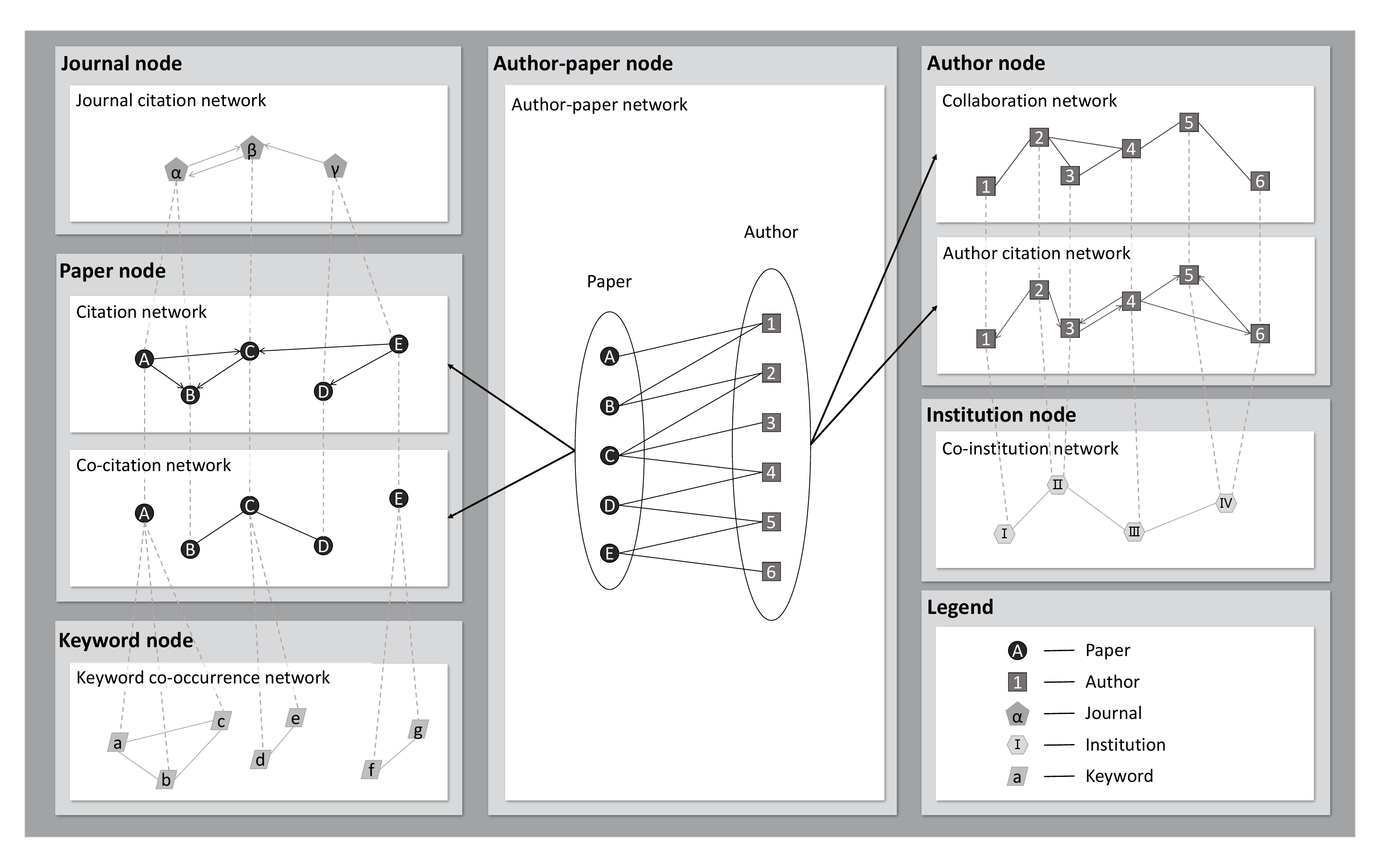}
        \caption{The structure of multi-layer academic networks.}
	\label{fig:network_structure}
\end{figure} 

\subsection*{Author and Paper Identification}

To derive the multi-layer academic networks from the publication dataset, one crucial challenge is author ambiguity, which can be classified into four main types. First, a single author may use various name formats in different articles, such as full names and initials. For example, Professor ``Wang Hansheng'' can be found in different formats, like ``Wang, H.'', ``Wang, Hansheng'' and ``Wang, HS''. Second, a name abbreviation may correspond to different authors. For example, ``Ding, Y'' might refer to ``Ding, Yang,'' ``Ding, Ying'' or ``Ding, Yu''. Third, an author's institutions may change over time. For instance, ``Wang, Hansheng'' has worked at different institutions such as ``University of Wisconsin-Madison'' and ``Peking University'' at different time periods. Last, the same name may refer to different authors. For instance, ``Li, Wei'' is a common name, and we have found six different ``Li, Wei'' authors affiliated with Delft University of Technology, Harbin Institute of Technology, Peking University, Renmin University of China, Syracuse University, and the University of Toledo, respectively. Failure to disambiguate author names could lead to inaccuracies in the underlying collaboration network, resulting in different authors being considered as the same person or the same person being identified as different authors. 

To address this issue, we implement a four-step process to identify authors, which corresponds to the left part in the ``Author and Paper Identification'' panel in Figure \ref{fig:data_collection}. First, we categorize authors by their initials and group those with the same initials together, treating those in different groups as different authors. This is just a preliminary classification; if the abbreviations of two authors are different, the two authors are most likely not the same person. Next, we use names and institutions to further distinguish authors within the same group. We measure the similarity between two strings based on the proportion of matching subsequences, with different standards for name and institution similarity. Specifically, we consider two authors' names as similar if their similarity score is no less than 0.8. Additionally, we consider institutions as similar if there is an intersection between them or their similarity score is no less than 0.9. We choose 0.9 because most of the names of institutions are standardized and uniform but there may exist a few spelling errors. Two authors are regarded as the same only if their names and institutions are both similar. Third, we may encounter situations where authors $i_1$ and $i_2$ are considered the same person and authors $i_1$ and $i_3$ are also considered the same person, but $i_2$ and $i_3$ are not. This often occurs when author $i_1$'s institutions are universities $l_1$ and $l_2$; author $i_2$'s institution is university $l_1$; and author $i_3$'s institution is university $l_2$. To address such scenarios, we adopt the concept of connected components in a graph and reconsider that authors $i_2$ and $i_3$ are the same person \cite{bondy1976graph}. Last, we manually check a portion of the authors, particularly those with common surnames like ``Chen,'' ``Li,'' ``Wang,'' ``Yang,'' ``Zhang,'' and many others. After name disambiguation, we identify 70,735 distinct authors and assign an {\it author unique id} to each of them. We would like to emphasize that managing author ambiguity is a challenging task and can be regarded as a contribution of our work.

In addition to the author identification problem, paper matching is also complicated. We cannot solely rely on the paper title to uniquely identify the paper, as there are cases where different papers share the same title but have different authors, publication time, and journals. For example, ``Linear model selection by cross-validation'' is the title of both a paper published in the {\it Journal of the American Statistical Association} by Professor Shao. J in 1993, and a separate paper published in the {\it Journal of Statistical Planning and Inference} by Professor Rao, CR and Wu, Y in 2005. Although such cases are relatively rare, accounting for only 0.0985\% of the total number of papers, they cannot be ignored. To address this problem, we employ a two-step method for paper identification and reference list matching. In the first step, we eliminate duplicate papers from the collected data by comparing the title and author information of each paper. We then assign a {\it paper unique id} to each paper. In the second step, we utilize the unique paper id to match the reference lists of papers, considering the title and author information. By incorporating author information into the process, we can effectively resolve the issue of paper citation matching.

\subsection*{Author attributes}

Four attributes of authors can be extracted from our dataset, namely research interest, productivity, region, and institution, corresponding to the panel of ``Author Attributes'' in Figure \ref{fig:data_collection}. We will provide a comprehensive explanation of how these attributes are processed. The first one is research interest.
To extract the research interest of each author, we employ the latent Dirichlet allocation (LDA) \cite{blei2003latent} method based on the titles of papers. Specifically, we select 30 topics from all the paper titles. Let $M$ be the number of titles (i.e., papers) and $P$ be the number of topics. Then, we construct a title-topic matrix denoted by $\mathbf{R} = (r_{jp}) \in \mathbb{R}^{M \times P}$, where $R_{jp}$ refers to the probability of the $p$th topic in the $j$th paper. To estimate the levels of interest of authors in these 30 topics, we gather the titles of the papers they have published and calculate the average probabilities for each topic based on $\mathbf{R}$. For instance, if an author $i$ has published papers $j_1$, $j_2$, and $j_3$, the estimated probability of the $p$th topic for that author is $q_{ip} = (r_{j_1p}+r_{j_2p}+r_{j_3p})/3$. Consequently, we can transform the matrix $\mathbf{R}$ to a matrix $\mathbf{Q} = (q_{ip}) \in \mathbb{R}^{N \times P}$, where $N$ is the number of authors and $q_{ip}$ represents the degree of interest of author $i$ towards the $p$th topic. The 30 topics can be interpreted based on their respective keywords, including ``survival analysis'', ``time series'', ``variable selection'' and others. The second attribute is productivity, which is a categorical attribute. For each author, we determine the value of this attribute based on the number of papers they have published in our dataset. Specifically, if an author has published no more than 5 papers, that author is labelled ``low productivity''. If an author has published more than 5 papers but no more than 10 papers, that author is labelled ``low-intermediate productivity''. If an author has published more than 10 papers but no more than 20 papers, they are labelled ``high-intermediate productivity''. If an author has published more than 20 papers, that author is labelled ``high productivity''. The third and fourth attributes are institution and region. For each author, we gather information on their institutions and the corresponding regions from ``Author Information'' in the panel of ``Paper Information.'' We identify the most frequently mentioned institution and region for each author as their institution and region. There are 980 (account for 1.39 \%) and 5,079 (account for 7.18 \%) authors who have no region and institution respectively, so their institutions and regions are recorded as ``Unknown''.

\subsection*{Construction of Multi-layer Academic Networks}

In this subsection, we provide a comprehensive description of multi-layer academic networks, including collaboration, co-institution, citation, co-citation, journal citation, author citation, author-paper, and keyword co-occurrence networks. The inner correlation of these networks is illustrated in Figure \ref{fig:network_structure}. Some of the networks are characterized by their weighted and dynamic nature, which contributes to the richness of our dataset.

\subsubsection*{Collaboration network} 

To construct the collaboration network, we first represent each author as a node and each collaboration between authors as an edge. If authors $i_1$ and $i_2$ have coauthored a paper, we define the corresponding entry in the $N\times N$ adjacency matrix $\mathbf{A}$ as $a_{i_1i_2}=1$. Otherwise, $a_{i_1i_2}=0$. Following convention, we set $a_{ii}=0$ for $i=1,\cdots,N$, where $N$ is the total number of authors. It is worth noting that the collaboration network is undirected, which implies that $a_{i_1i_2} = a_{i_2i_1}$. We can also create a weighted collaboration network by defining the weighted adjacency matrix $\mathbf{W} = (w_{i_1i_2})\in\mathbb{R}^{N\times N}$, where $w_{i_1i_2}$ denotes the frequency with which authors $i_1$ and $i_2$ have coauthored papers. In addition, we provide the dynamic collaboration networks that span over the years, which can be used for link prediction on collaboration patterns. 

\subsubsection*{Co-institution network} 

A co-institution network is a type of network where nodes represent institutions, such as universities, and the links represent collaborations between them. In other words, it is a network of institutions that work together on published papers. If an author from institution $l_1$ coauthored with an author from institution $l_2$, then an edge exists between $l_1$ and $l_2$. The adjacency matrix of this network is defined as $\mathbf{D} = (d_{l_1l_2})\in\mathbb{R}^{L\times L}$, where $L$ is the number of institutions in the network. Note that this network carries weight, representing the frequency that authors from two institutions collaborate. Additionally, for any $1\leq l\leq L$, the diagonal element $d_{ll}$ represents the frequency that authors collaborate within the same institution. 

\subsubsection*{Citation network} 

The citation network is constructed by examining the reference list of each paper. Specifically, if paper $j_1$ includes a reference to paper $j_2$, then we denote this relationship as $b_{j_1j_2}=1$, otherwise $b_{j_1j_2}=0$. The resulting adjacency matrix of the citation network is denoted as $\mathbf{B} = (b_{j_1j_2})\in\mathbb{R}^{M\times M}$, where $M$ is the total number of papers in the network. Since a paper cannot cite itself, we always let $b_{jj}=0$ for $j = 1,\cdots, M$. Note that the citation network is directed, meaning that if paper $j_1$ cites $j_2$, paper $j_2$ will not cite paper $j_1$. To be more precise, the citation network is a directed acyclic graph (DAG), with no cycles present in the network. We also establish the dynamic citation network, recording by a series of the adjacency matrices $\mathbf{B}_t = (b_{j_1j_2}^{(t)})\in\mathbb{R}^{M_t\times M_t}$ with $t = 1,\cdots, T$, where $M_t$ is the total number of papers in the network at time $t$. A specific $\mathbf{B}_t$ captures the citation relationships within a specific time period. 

\subsubsection*{Co-citation network} 

The co-citation network is constructed from the citation network, which reveals the co-citation patterns among papers. When two papers $j_1$ and $j_2$ are both cited by the same third paper $j$, an edge is created between $j_1$ and $j_2$, and this relationship is recorded as $c_{j_1j_2}=1$. Otherwise, $c_{j_1j_2}=0$. Additionally, set $c_{jj}=0$. As a result, the adjacency matrix of the resulting co-citation network can be represented as $\mathbf{C}=(c_{j_1j_2})\in\mathbb{R}^{M\times M}$. The co-citation network is undirected as co-citation is a symmetric relationship. We also construct a weighted co-citation network in which the weight of the edge is equal to the number of other papers that cite both.

\subsubsection*{Journal citation network} 

A journal citation network is a network that illustrates the relationships among academic journals based on the citations they make to each other. This network is composed of nodes that represent journals and edges that signify the citations between them. To be specific, if a paper published in journal $k_1$ cites a paper published in journal $k_2$, then an edge exists from node $k_1$ to node $k_2$ in the network. This relationship is captured by an adjacency matrix $\mathbf{S}=(s_{k_1k_2})\in\mathbb{R}^{K\times K}$, where $s_{k_1k_2}=1$ represents the existence of the citation relationship while $s_{k_1k_2}=0$ represents the absence of a citation and $K$ is the total number of journals. The journal citation network is directed, and the edges have weights that correspond to the number of papers that cite journal $k_1$ in journal $k_2$. The diagonal elements represent the self-citation behavior within the same journal. 

\subsubsection*{Author citation network}

The author citation network is a representation of the relationships among authors based on the citations they receive in academic publications. If a paper published by author $i_1$ cites a paper published by author $i_2$, the relationship can be denoted as $e_{i_1i_2}=1$, otherwise $e_{i_1i_2}=0$. The adjacency matrix of the resulting author citation network can be represented as $\mathbf{E}=(e_{i_1i_2})\in\mathbb{R}^{N\times N}$. The author citation network is directed and 
it allows for the presence of mutual citation relationships between nodes, which is different from the citation network. Additionally, different from the collaboration network, the diagonal elements of the author citation network are allowed to be non-zero, reflecting the self-citation behavior of authors. We also provide a weighted author-citation network where the edge weights represent the number of citations.

\subsubsection*{Author-paper network} 

The author-paper network is a type of network that represents the relationship between authors and their published papers. This type of network is a two-mode network, which means that it consists of two types of nodes, namely, authors and papers. The network is constructed by establishing edges between authors and papers. Specifically, the relationship is represented by the adjacency matrix $\mathbf{H} = (h_{ij})\in\mathbb{R}^{N\times M}$. The elements of the adjacency matrix are defined such that $h_{ij} =1$ if author $i$ has written paper $j$, and $h_{ij} = 0$ otherwise.  

\subsubsection*{Keyword co-occurrence network} 

Due to the limitations of the original publication data, we cannot obtain the keywords of papers published before 1992. Therefore, the keyword co-occurrence network is constructed from published articles during 1992--2021, where the node represents the keyword.
Due to the large variety of keyword types and formats, we have filtered out keywords that appear ten times or more and conducted a simple cleaning process on them.
If two keywords appear together in one article, there exists an edge between them. The keyword co-occurrence network is undirected and carries weight, that is, the frequency with which two keywords co-exist. It is believed that the keyword co-occurrence network is important for capturing the knowledge structure in a certain research area \cite{cheng2020keyword}.

\section*{Data Records}

The LMANStat dataset is publicly available on GitHub, and can be accessed directly at \href{https://github.com/Gaotianchen97/LMANStat}{github.com/Gaotianchen97/LMANStat}. The dataset comprises two Excel files that contain the author and paper information (including author unique id, paper unique id, and the corresponding attributes) and eight CSV files that depict the edge list. Specifically, the record of the papers can be found in the file {\it Paper\_information.xlsx}, which contains the basic information of the papers as provided in Table \ref{tab:data_example}. Each row of the table represents a unique academic paper and is identified by the unique field {\it Paper\_unique\_id}. The file {\it Author\_information.xlsx} contains the attributes of each unique author, including the research interest, productivity, institution, and region, and authors can be uniquely identified by the field {\it Author\_unique\_id}. The eight CSV files with the name {\it Edgelist\_X.csv} (X is the name of each network) correspond to the edge lists of the eight networks described above. The fields {\it (Target and Source)} in each row of the edge data files represent an edge in the corresponding network, with the field {\it Year} indicating the time in which the edge is formed. By filtering according to the {\it Year} field, one can easily construct the academic network dynamically. 
The nodes in files {\it Edgelist\_citation.csv}, {\it Edgelist\_co\_citation.csv}, {\it Edgelist\_collaboration.csv}, {\it Edgelist\_author\_citation.csv}, and {\it Edgelist\_author\_paper.csv} are {\it Paper\_unique\_id} or {\it Author\_unique\_id}, while the nodes in files {\it Edgelist\_journal\_citation.csv}, {\it Edgelist\_co\_institution.csv}, and {\it Edgelist\_keyword\_co\_occurrence.csv} are the journal names, institution names, and keywords, respectively. Table \ref{tab:intro_network} provides a comprehensive overview of our multi-layer networks, including the network name, node and edge definitions, the total number of nodes and edges, and the characteristics of networks (U-Undirected, D-Directed, W-Weighted, B-Bipartite). It is worth noting that our multi-layer networks are notably large in scale and dynamic in nature. 

\begin{table}[ht]
\small
\centering
\newcommand{\tabincell}[2]{\begin{tabular}{@{}#1@{}}#2\end{tabular}}
\begin{tabular}{cccccc}
    \hline
    \textbf{Network} & \textbf{Node} & \textbf{\# of nodes} & \textbf{Edge} & \textbf{\# of edges} & \textbf{Characteristics}\\
    \hline
    Collaboration & Author & 67,353 & Co-authorship & 197,171 & U \& W\\
    Co-institution & Institution & 3,542 & Co-authorship & 110,017 & U \& W\\
    Citation & Paper & 83,996 & Citation relationship & 517,079 & D\\
    Journal citation & Journal & 42 & Citation relationship & 28,830 & D \& W\\
    Author citation & Author & 59,334 & Citation relationship & 2,083,857 & D \& W \\
    Co-citation & Paper & 58,617 & Co-citation relationship & 2,956,432 & U \& W \\
    Author-paper & Author \& Paper & 70,735 \& 97,436 & Publication relationship & 217,774 & B\\
    Keyword co-occurrence & Keyword & 5,037 & Co-occurrence relationship & 374,714 & U \& W \\
    \hline
    \end{tabular}
    \caption{\label{tab:intro_network} The brief description of the multi-layer academic networks.}
\end{table}

\section*{Technical Validation}

In this section, we aim to validate the dataset through various potential scenarios for exploring and analyzing our multi-layer academic networks. To emphasize the usability of our dataset, key insights into the characteristics of the data are also provided, aligning with historical research findings and the consensus among statisticians. More importantly, the LMANStat dataset is extensively utilized by our research team to validate its usability \cite{zhang2023community,song2022link,gao2021community,gao2023citation}. In our multi-layer academic networks, the collaboration network and citation network are the most commonly used networks. Therefore, we utilize them for verification. Additionally, we also consider the journal citation network with journals as nodes and the keyword co-occurrence network with keywords as nodes to validate the LMANStat dataset.

 \subsection*{Validation of the Collaboration and Citation Network}

We first focus on the most extensively studied academic networks, i.e., the collaboration and the citation network. To be specific, we identify $N = 67,353$ unique authors in the collaboration network and $M = 83,996$ papers in the citation network, leading to a density of 0.00869\% and 0.00733\%, respectively. Note that the network density is extremely low, which is consistent with the existing research findings \cite{jin2015fast,ji2022co}. It should be noted that the total number of authors we identified exceeds the number of authors in the collaboration network, as some papers are completed by independent authors who do not participate in any collaborations in our dataset. 

\begin{figure}[!htp]
    \centering    \includegraphics[width=0.90\textwidth]{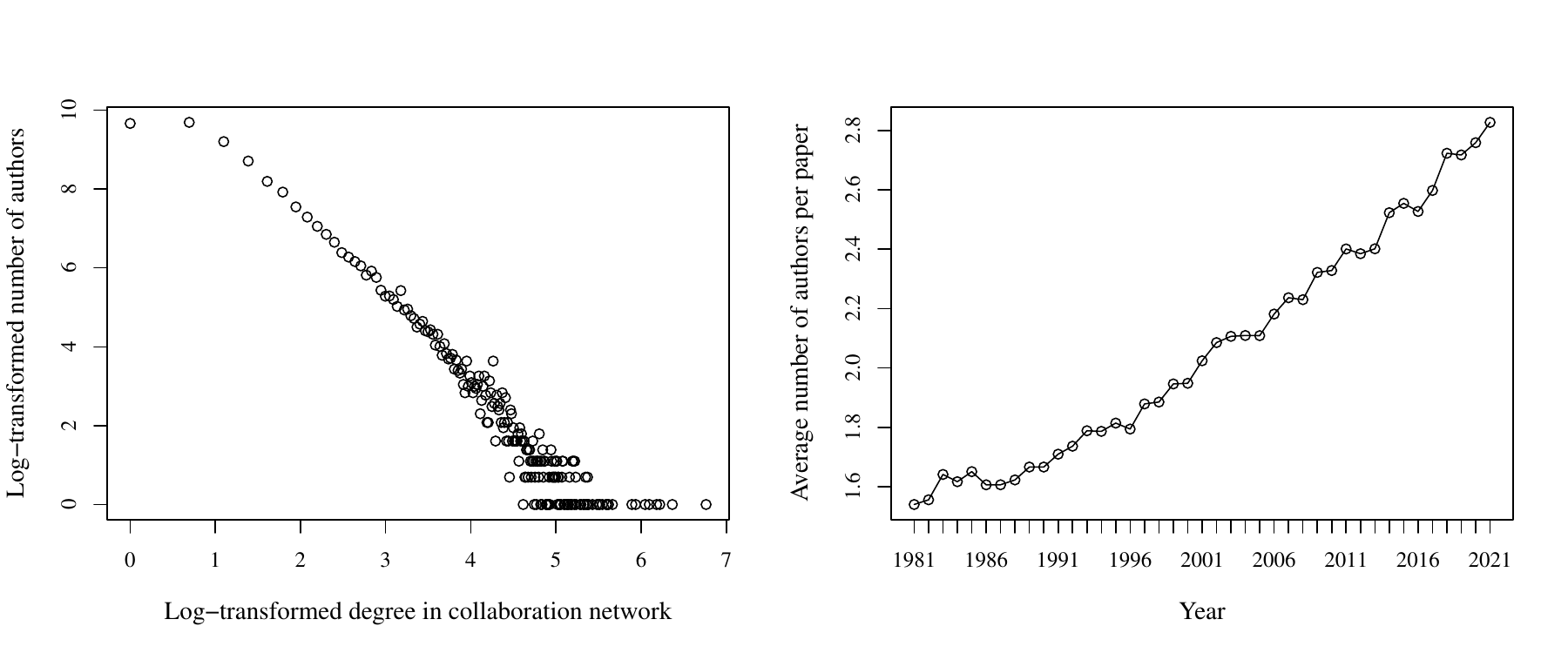}
    \caption{Left: Log-log degree distribution in collaboration network; Right: Average number of authors per paper per year.}
    \label{fig:Loglog_author_num}
\end{figure}

We define the degree of author $i$ as $\mbox{deg}_i = \sum_{i'\not=i}a_{ii'}$; a higher degree indicates that an author has more collaborators. Furthermore, we present a plot of the log-log degree distribution in the left subplot in Figure \ref{fig:Loglog_author_num}, where the scale-free phenomenon can be detected \cite{barabasi1999emergence}. This phenomenon is also referenced in research across collaborative networks \cite{yan2010mapping,ji2022co}. The time series of average number of authors per paper is also reported in the right subplot in Figure \ref{fig:Loglog_author_num}. It can be seen that the average number of authors per paper shows an increasing trend by year, indicating the phenomenon of collaboration on research in the field of statistics. It is also validated in the literature that extensive collaboration is detected in various research fields \cite{ji2016coauthorship,ji2022co}.

\begin{figure}[h]
    \centering
    \includegraphics[width=0.50\textwidth]{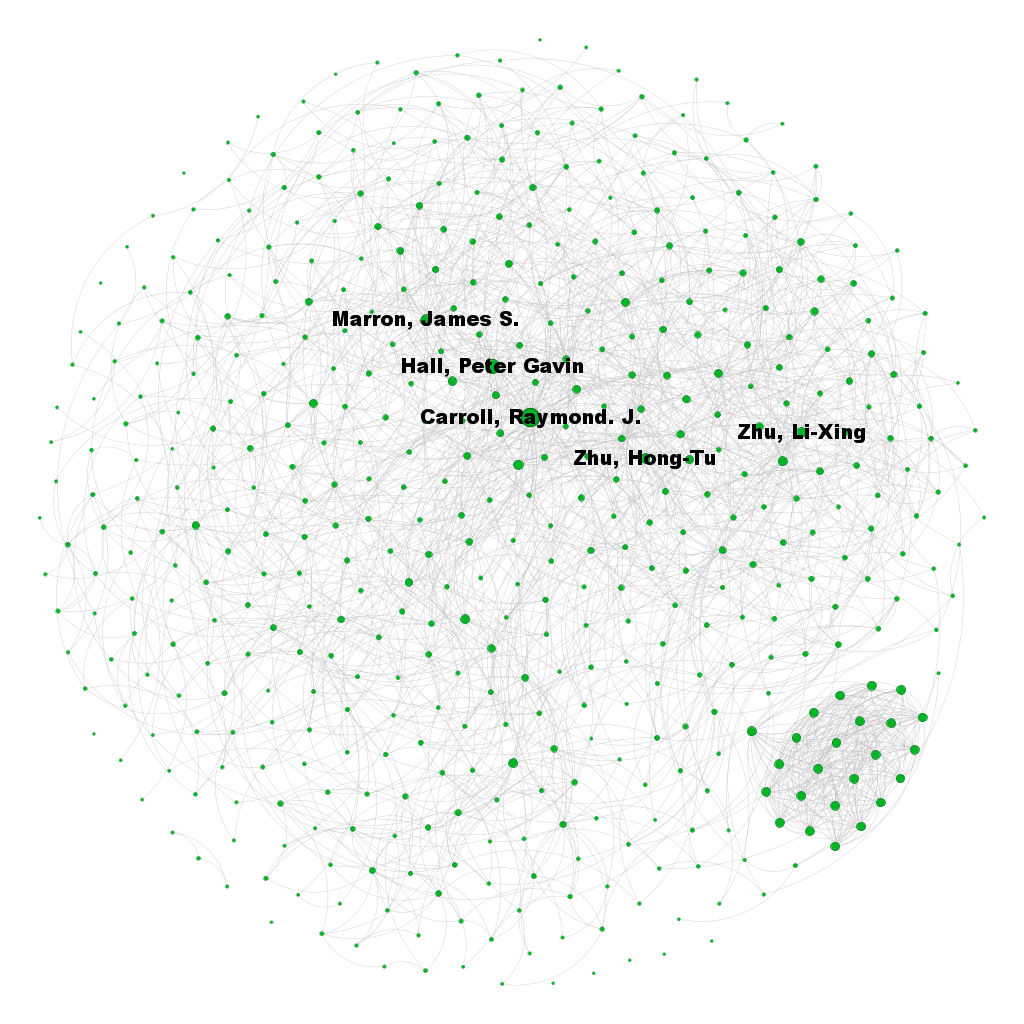}
    \caption{Collaboration network of statisticians whose degrees are no less than 40. The greater the degree of a node, the greater the size of the node. }
    \label{fig:collaboration_network}
\end{figure}

Figure \ref{fig:collaboration_network} shows a sub-network of the collaboration network. It is composed of authors with degree larger than or equal to 40 in the original network, leading to 487 authors and 2,036 edges. Regarding the collaboration network, the {\it clustering coefficient} indicates the likelihood of collaboration between two authors when they jointly work with a third author. The clustering coefficient in the above collaboration network is 0.368. The result aligns with the findings from other literature, which states that the clustering coefficients of the theoretical physics networks ranges from 0.327 to 0.430 \cite{newman2001structure}. Additionally, the names of the top five authors holding the highest degrees are clearly indicated in Figure \ref{fig:collaboration_network}.  Their innovative approaches and insights generate a lasting impact in the field of statistics. For instance, Professor Carroll Raymond J. has made substantial contributions to various domains within statistical methodology and theory. The applications of his research extend across diverse fields, including radiation and nutritional epidemiology, molecular biology, genomics, and numerous others. To conclude, the presented evidence indicates the exceptional quality of our collaboration network.

As for the citation network, the in-degree of a paper is crucial as it represents the number of times the paper has been cited within the network. We define the in-degree of paper $j$ as $\mbox{deg}_{+j}=\sum_{j'\not=j}b_{j'j}$; a higher in-degree implies a greater number of citations of a paper. Within our dataset, the average in-degree per paper is 5.31, which is closely related to the Impact Factor (IF) of the chosen journals. In our LMANStat dataset, 37,592 papers (38.5\%) have not received citations from any other paper, 24,162 papers (24.8\%) do not cite any other papers, and 13,440 papers (13.8\%) neither cite nor are cited by any other papers. This quantitative distribution aligns with findings in previous research \cite{ji2016coauthorship}. The {\it Gini coefficient}, calculated at 0.76, indicates a substantial dispersion in the in-degree values. This result parallels the Gini coefficient (0.77) observed in the other citation network \cite{ji2016coauthorship}, providing further affirmation of the credibility and quality of our citation network.

Subsequently, we proceed to assess the rationality of important nodes within the citation network. We are particularly interested in {\it citation counts}, as reported by the Web of Science. It is believed that papers with high citation counts are considerably influential, as they have made substantial contributions to their respective fields \cite{bornmann2012factors}. Table \ref{tab:paper_large_degree} lists the top 10 papers with the largest citation counts in our citation network. These papers hold key positions within the field of statistics, covering a diverse array of research directions. These include topics such as variable selection, multiple testing, nonparametric statistics, causal inference, regression analysis, machine learning, and more. To illustrate, the first paper introduced the concept of the false discovery rate (FDR) and proved a simple sequential Bonferroni-type procedure to control the FDR for independent test statistics \cite{benjamini1995controlling}. This exemplifies the influential nature of the research discussed. 

\begin{table}[!ht]
\centering
\newcommand{\tabincell}[2]{\begin{tabular}{@{}#1@{}}#2\end{tabular}}
\begin{tabular}{ccccc}
    \hline
    \textbf{ID} & \textbf{Title} & \textbf{Journal} & \textbf{\tabincell{c}{Published\\ date}} & \textbf{\tabincell{c}{Citation\\ counts}}\\
    \hline
    1 & \tabincell{c}{Controlling the false discovery rate - a practical and powerful \\approach to multiple testing} & JRSS-B & 1995 & 57,116 \\
    2 & Fitting linear mixed-effects models using lme4 & JSS & 2015 & 31,080 \\
    3 & Regression shrinkage and selection via the Lasso & JRSS-B & 1996 & 21,383 \\
    4 & \tabincell{c}{The central role of the propensity score in observational\\ studies for causal effects} & Biometrika & 1983 & 14,825 \\  
    5 & \tabincell{c}{Longitudinal data-analysis using generalized linear-models} & Biometrika & 1986 & 12,824 \\
    6 & \tabincell{c}{Comparing the areas under 2 or more correlated receiver \\operating characteristic curves - a nonparametric approach} & Biometrics & 1988 & 12,699 \\
    7 & \tabincell{c}{Operating characteristics of a bank correlation test for\\ publication bias} & Biometrics & 1994 & 10,936 \\
    8 & \tabincell{c}{A proportional hazards model for the subdistribution of \\a competing risk} & JASA & 1999 & 8,265 \\
    9 & \tabincell{c}{Greedy function approximation: a gradient boosting machine} & AoS & 2001 & 8,262 \\
    10 & \tabincell{c}{Exploration, normalization, and summaries of high density\\ oligonucleotide array probe level data} & Biometrics & 2003 & 8,169 \\
    \hline
    \end{tabular}
    \caption{\label{tab:paper_large_degree} The top 10 papers with the largest citation counts. Paper title, journal name, published date, and citation counts are reported.}
\end{table}

\subsection*{Validation of the Journal Citation Network}

Journal citation networks are often employed for ranking journals, which is considered an important indicator for evaluating the quality and impact of publications in specific research fields. Therefore, we validate the accessibility of the journal citation network through journal ranking.
Nowadays, numerous journal ranking indicators are available, including the Impact Factor (IF), SCImago Journal Rank (SJR), CiteScore, and others. These indicators employ diverse metrics and methodologies to assess and rank journals based on various factors such as citation counts, publication frequency, editorial standards, and scholarly influence. 
In our study, we attempt to rank journals using the PageRank centrality derived from the journal citation network. Following tradition \cite{heneberg2016excessive}, we ignore self-citations among journals. By calculating the PageRank centrality of each node (journal), we can effectively rank journals based on their importance. Interestingly, we observe the phenomenon that the ranking of journals based on PageRank centrality closely aligns with the expectations and intuitions of statisticians. This suggests that the PageRank-based approach provides a ranking that resonates well with the perceptions of experts in the field.

\begin{figure}[!htp]
    \centering
    \includegraphics[width=0.6\textwidth]{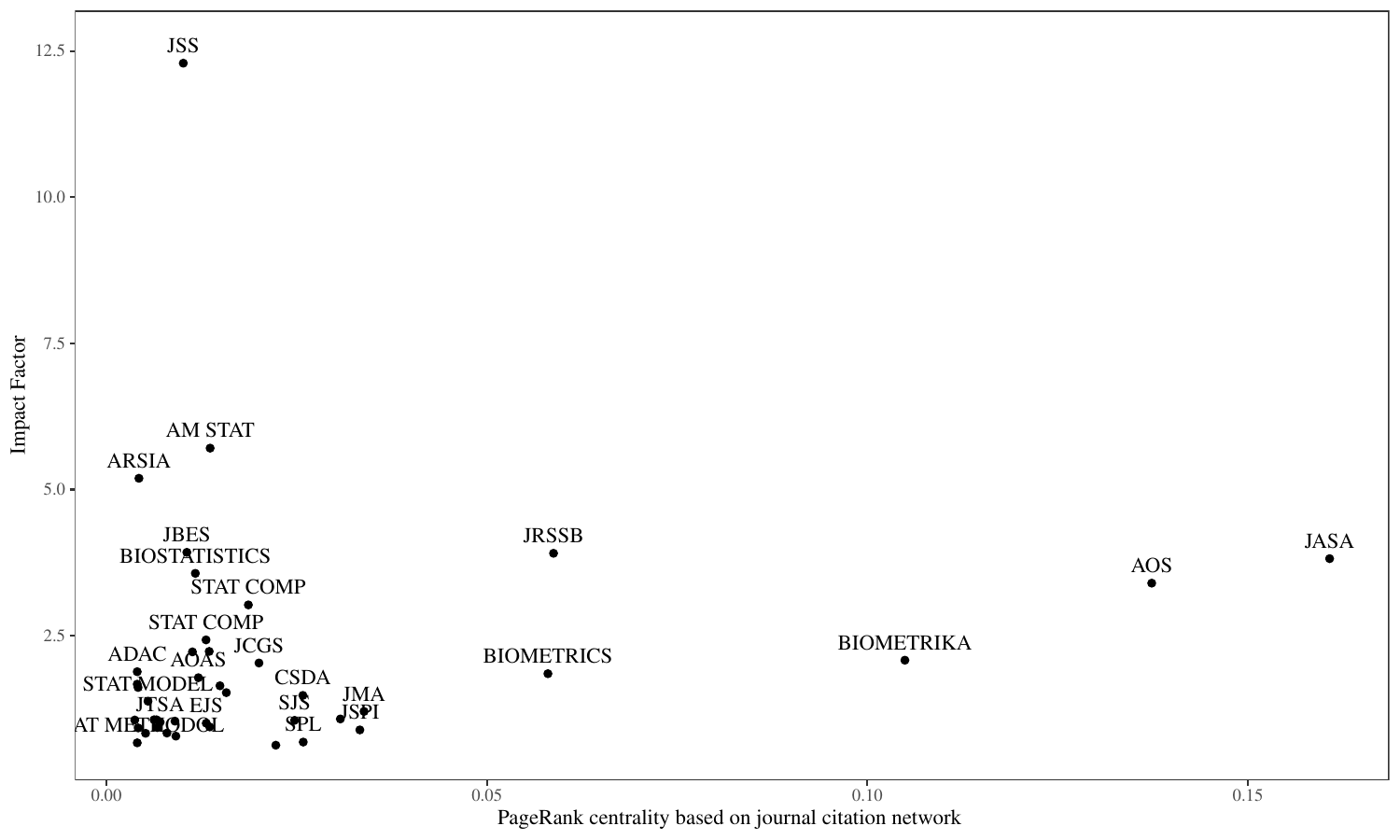}
    \caption{Scatter plot of PageRank centrality and Impact Factor (IF).}
    \label{fig:pagerank_if}
\end{figure}

Figure \ref{fig:pagerank_if} shows a scatter plot that compares journal PageRank centrality and IF. Notably, the highest-ranked statistical journals based on IF include the JSS ({\it Journal of Statistical Software}), AM STAT ({\it American Statistician}), and ARSIA ({\it Annual Review of Statistics and Its Application}), all of which hold important influence in the field. It is worth mentioning that JSS is an influential journal that is dedicated to the development and application of software and computational methods, while the latter two are review-oriented publications. Regarding the x-axis, which represents PageRank centrality based on the journal citation network. To be specific, {\it Biometrika} and {\it Biometrics} are well-recognized journals in the fields of biostatistics and biometrics.  AoS ({\it Annals of Statistics}), JASA ({\it Journal of the American Statistical Association}), and JRSS-B ({\it Journal of the Royal Statistical Society Series B-Statistical Methodology}) are highly respected statistical journals that are dedicated to statistical methodology and theory, as well as applications in various disciplines. Additionally, the AoS, JASA, JRSS-B, and Biometrika are regarded as the top four journals by statisticians. 
This phenomenon precisely validates the advantage of the journal citation network in our LMANStat dataset. 

\subsection*{Validation of the Keyword Co-occurrence Network}

We conduct quality validation on the keyword co-occurrence network in this subsection. In LMANStat dataset, the keyword co-occurrence network contains 5,037 keywords, which cover the most popular research areas in statistics. 
Table \ref{tab:top20nodepair} lists the 20 most frequently appeared keyword pairs. The first pair, ``asymptotic normality and consistency,'' represents the most prevalent property of estimators in statistics. The second pair, ``Bayesian inference and MCMC,'' commonly refers to the use of MCMC method for parameter estimation and posterior inference in Bayesian inference. The third pair, ``Gibbs sampling and MCMC,'' denotes the utilization of Gibbs sampling as a specific implementation of the MCMC for sampling from multivariate probability distributions. Furthermore, there are several other frequently observed keyword pairs. For instance, ``LASSO and variable selection'' frequently co-occur, since LASSO is a popular method for variable selection. ``Kurtosis and skewness'' represent two statistical measures that provide insights into the shape and distribution of a dataset. ``Akaike information criterion and Bayesian information criterion'' are both statistical criteria employed for model selection and comparison. The keyword co-occurrence network constructed in our dataset encompasses the vast majority of research directions within the field of statistics. It serves as fertile ground for studying the development and evolution of statistical subject matters. In summary, the above phenomenon demonstrates that the network aligns with the intuition of statisticians and affirms the high quality of the dataset.

\begin{table}[h]
\centering
\begin{tabular}{cccccc}
\hline
ID & Keyword Pair  & Frequency & ID & Keyword Pair  & Frequency  \\ \hline
1  & asymptotic normality \& consistency  & 249       & 11 & kurtosis \& skewness  & 117 \\
2  & {\begin{tabular}[c]{@{}c@{}}Bayesian inference \& \\ Markov chain Monte Carlo\end{tabular}} & 249   & 12 & {\begin{tabular}[c]{@{}c@{}}Markov chain Monte Carlo \& \\ Metropolis Hastings\end{tabular}}  & 112  \\
3  & {\begin{tabular}[c]{@{}c@{}}Gibbs sampling \& \\ Markov chain Monte Carlo\end{tabular}} & 216       & 13 & {\begin{tabular}[c]{@{}c@{}}average run length \& \\ statistical process control \end{tabular}}   & 109  \\
4  & bias \& mean squared error  & 189   & 14 & false discovery rate \& multiple testing  & 108   \\
5  & {\begin{tabular}[c]{@{}c@{}}expectation maximization algorithm \& \\ maximum likelihood estimation\end{tabular}} & 157  & 15 & {\begin{tabular}[c]{@{}c@{}}Akaike information criterion \& \\ Bayesian information criterion \end{tabular}} & 103  \\
6  & {\begin{tabular}[c]{@{}c@{}}Bayesian estimation \& \\ maximum likelihood estimation\end{tabular}}    & 150 & 16 & {\begin{tabular}[c]{@{}c@{}}generalized estimating equation \& \\ longitudinal data\end{tabular}}   & 102 \\
7  & sensitivity \& specificity & 138  & 17 & Gibbs sampling \& Metropolis Hastings & 94  \\
8  & {\begin{tabular}[c]{@{}c@{}}expectation maximization algorithm \& \\ mixture model \end{tabular}}  & 137   & 18 & {\begin{tabular}[c]{@{}c@{}}hierarchical model \& \\ Markov chain Monte Carlo\end{tabular}}   & 93  \\
9  & Bayesian \& Markov chain Monte Carlo & 132       & 19 & {\begin{tabular}[c]{@{}c@{}}expectation maximization algorithm \& \\ missing data\end{tabular}}   & 91   \\
10 & lasso \& variable selection  & 125   & 20 & Bayesian inference \& Gibbs sampling   & 88 \\ \hline
\end{tabular}
\caption{Top 20 most frequent keyword pairs during 1992-2021.}
\label{tab:top20nodepair}
\end{table}

\section*{Usage Notes}

In this section, we focus on providing insights into the research possibilities and examples that can be conducted using the LMANStat dataset. We use the citation network as an example to illustrate the dynamic nature of multi-layer academic networks. Additionally, we present a selection of applications and examples.

\subsection*{Dynamic Nature of the Citation Network}

\begin{figure}[!ht]
\centering
\subfigure[Network from 1980 to 2006]{
\includegraphics[width=0.32\textwidth]{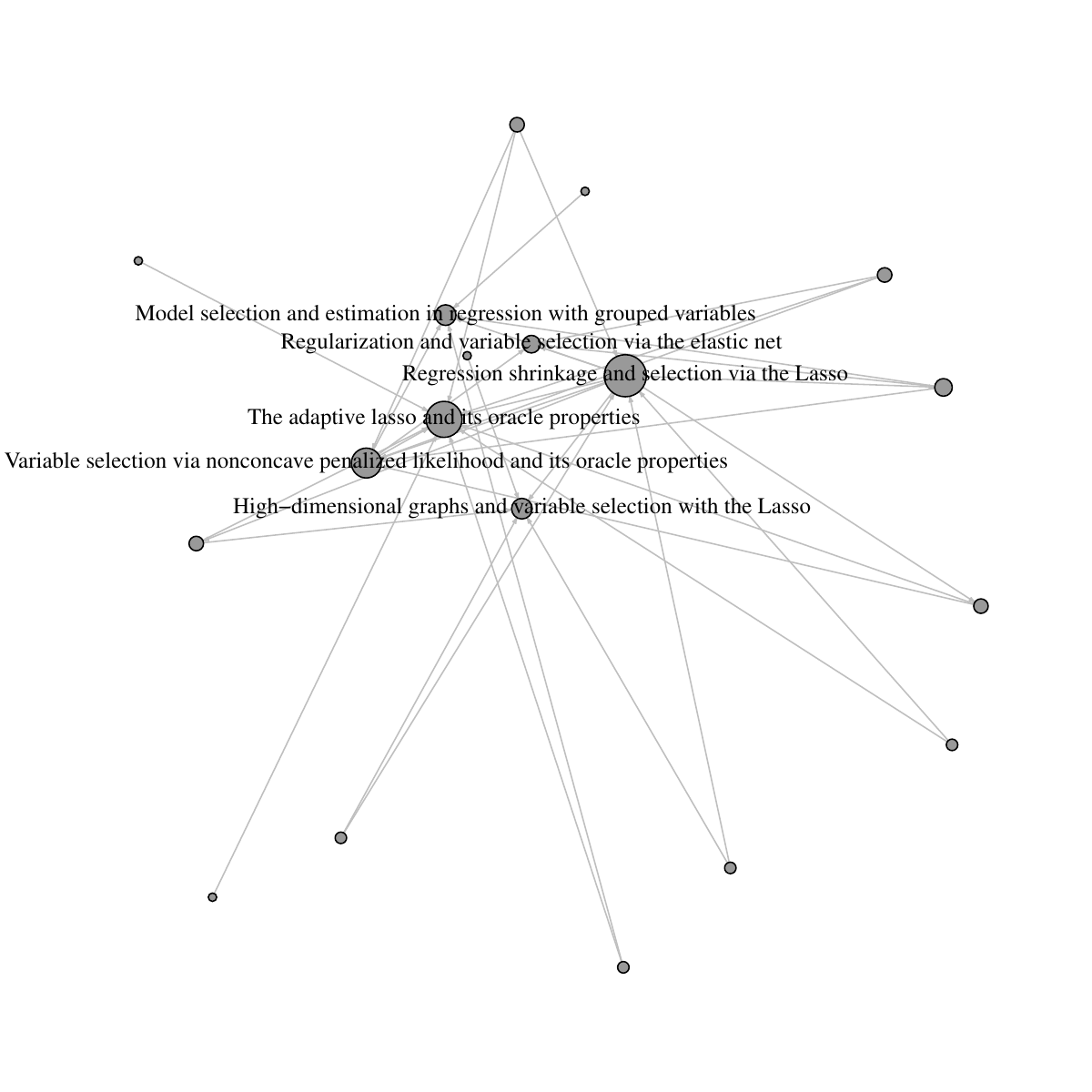}}
\subfigure[Network from 1980 to 2010]{
\includegraphics[width=0.32\textwidth]{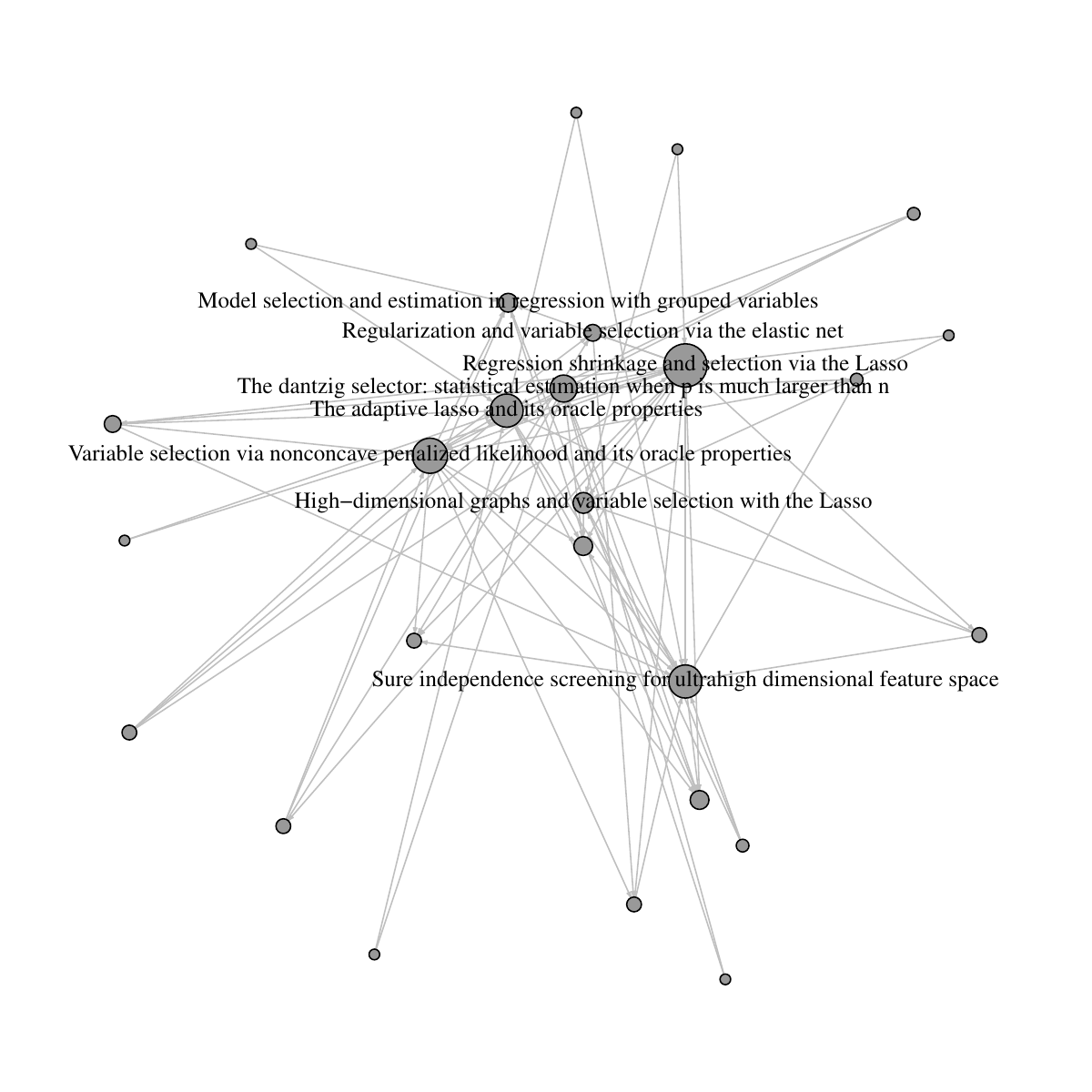}}
\subfigure[Network from 1980 to 2020]{
\includegraphics[width=0.32\textwidth]{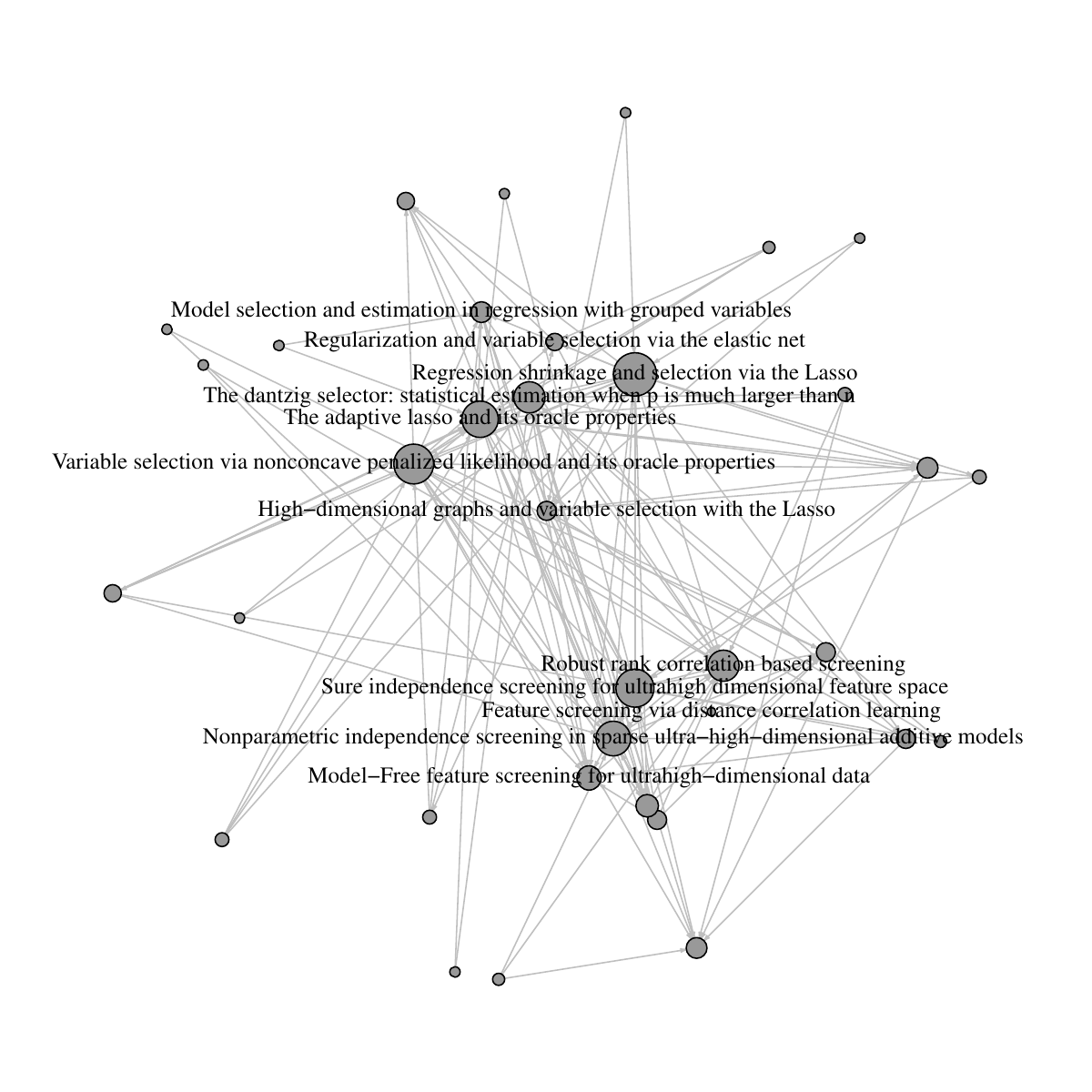}}
\caption{Visualization of citation networks composed of papers related to variable selection at different snapshots. It can be seen that the citation network has a community structure and that the communities are constantly evolving.}
\label{fig:visual}
\end{figure}

It is important to note that the multi-layer academic networks presented in this paper are all dynamic in nature. In this subsection, taking the citation network as an illustration, we showcase the dynamic nature
of the network. Figure \ref{fig:visual} provides a visualization of subnetworks within the citation network, focusing on papers related to ``variable selection''. This analysis aims to examine the  dynamic nature of the citation network over time. The visualization includes snapshots of the network from different time periods: 1980--2006, 1980--2010, and 1980--2020. It is evident from the visualization that the citation network exhibits a community structure that undergoes constant changes over time. This community shows continuous growth over the years, as indicated by the increasing number of papers associated with variable selection. Furthermore, as time progresses, the variable selection community gradually splits into two distinct communities. The first community focuses on traditional variable selection methods like the LASSO and other conventional techniques. The second community, referred to as the ultrahigh-dimensional community, explores variable selection techniques specifically tailored to challenges posed by the big data era. This community investigates methods such as screening-based approaches, which aim to address the unique requirements and complexities of analyzing large-scale datasets with numerous variables. In conclusion, it is worth emphasizing that the networks within this dataset are all dynamic, thereby enabling the exploration of dynamic nature.

\subsection*{Applications and Examples}

Based on the LMANStat datasets, researchers have the opportunity to address numerous research questions related to networks. These questions encompass link prediction, community detection, citation count prediction, and many others. For instance, meaningful communities (i.e., research areas) can be detected based on dynamic citation networks, and link prediction can be performed on the collaboration network for authorship recommendations. In Table \ref{Tab:application}, we present a selection of application examples. However, it is important to note that the exploration of this dataset extends far beyond these suggestions, with the potential to uncover interesting findings.

\begin{table}[!ht]
    \small
    \newcommand{\tabincell}[2]{\begin{tabular}{@{}#1@{}}#2\end{tabular}}
    \begin{center}
    \begin{tabular}{cl}
    \hline
    \hline
    \textbf{ID} & \textbf{Description}\\
    \hline
    1 & \tabincell{l}{\textbf{Title:} {\sc Community Detection for Statistical Citation Network by D-SCORE}\\ \textbf{Abstract:} [...]In this paper, we analyze \textbf{a citation network of the top 4 statistical journals from 2001 to 2018}, applying the \\directed spectral clustering on the ratio-of-eigenvectors (D-SCORE) method to detect the community structure of the citation\\ network. We find that[...]\\ \textbf{DOI:} \href{http://dx.doi.org/10.4310/20-SII636}{http://dx.doi.org/10.4310/20-SII636}}   \\
    \hline
    2 & \tabincell{l}{\textbf{Title:} {\sc Community Detection in Temporal Citation Network via a Tensor-based Approach}\\ \textbf{Abstract:} [...]Detecting and tracking community evolution in temporal networks can uncover important and interesting\\ behaviors. In this paper, we analyze \textbf{a temporal citation network constructed by publications collected from 44 statistical} \\ \textbf{journals between 2001 and 2018.} We propose[...]\\ \textbf{Web:} \href{https://www.intlpress.com/site/pub/pages/journals/items/sii/_home/acceptedpapers/index.php}{SII Accepted}}   \\
    \hline
    3 & \tabincell{l}{\textbf{Title:} {\sc Community Detection in Attributed Collaboration Network for Statisticians}\\ \textbf{Abstract:} [...] In this work, we collect \textbf{papers published between 2001 and 2018 in 43 statistical journals} and investigate\\ the collaborative trends and patterns. We find that more and more researchers take part in statistical research, and cooperation\\ among them is strengthening. We further construct \textbf{an attributed collaboration network and extract its core}[...]\\ \textbf{DOI:}  \href{http://dx.doi.org/10.1002/sta4.507}{http://dx.doi.org/10.1002/sta4.507}}   \\
    \hline
    4 & \tabincell{l}{\textbf{Title:} {\sc Link Prediction for Statistical Collaboration Networks Incorporating Institutes and}\\ {\sc Research Interests}\\ \textbf{Abstract:} [...] In this study, we construct \textbf{collaboration networks based on the co-authorship information of the papers}\\ \textbf{published in 43 statistical journals from 2001 to 2018.} We construct training and testing networks according to the timestamps\\ of the papers and construct a classification dataset for link prediction. We calculate[...]\\ \textbf{DOI:}  \href{http://dx.doi.org/10.1109/ACCESS.2022.3210129}{http://dx.doi.org/10.1109/ACCESS.2022.3210129}}   \\
    \hline
    5 & \tabincell{l}{\textbf{Title:} {\sc Citation Counts Prediction of Statistical Publications based on Multi-layer Academic Networks}\\ {\sc via Neural Network Model}\\ \textbf{Abstract:} [...]We collect \textbf{55,024 academic papers published in 43 statistics journals between 2001 and 2018.} Furthermore, \\we collect and clean a high-quality dataset and then construct multi-layer networks from different perspectives. Additionally,\\ we extract 77 factors for citation counts prediction, including 22 traditional and 55 network-related factors[...]\\ \textbf{DOI:} \href{http://dx.doi.org/10.2139/ssrn.4406701}{http://dx.doi.org/10.2139/ssrn.4406701}}   \\
    \hline
	\end{tabular}
	\end{center}
	\caption{\label{Tab:application} Application examples based on LMANStat dataset}
\end{table}

\section*{Code availability}

The code used for constructing the network and conducting preliminary exploration of the network can be accessed publicly on GitHub and is permanently available at \href{https://github.com/Gaotianchen97/LMANStat}{github.com/Gaotianchen97/LMANStat}. We provide both R and Python versions of the code. The Python code is written in Python 3.7 and utilizes version 2.6.3 of the ``networkx'' package. The R code is written in R 4.1.1 and uses version 1.2.6 of the ``igraph'' package. Both the R and Python codes use a citation network as an example to demonstrate network construction, basic descriptive analysis, and visualization, which contains detailed comments and usage recommendations to facilitate code reuse. 

\bibliography{sample}


\section*{Acknowledgements} 

The research of Rui Pan is supported by National Natural Science Foundation of China (No, 11971504), the Disciplinary Funds and the Emerging Interdisciplinary Project of Central University of Finance and Economics. The research of Rui Pan is also supported by the Program for Innovation Research in Central University of Finance and Economics.
The research of Hansheng Wang is partially supported by National Natural Science Foundation of China (No, 12271012). 

\section*{Author contributions statement}

T.G. and Y.Z. collected and cleaned the dataset. T.G., Y.Z., and R.P. conducted data analysis and visualized the results. R.P. and H.W. organized the paper. All authors wrote and reviewed the manuscript. 

\section*{Competing interests}

The authors declare no competing interests.

\end{document}